\documentclass[letterpaper]{article} 
\usepackage{aaai2026}  
\usepackage{times}  
\usepackage{helvet}  
\usepackage{courier}  
\usepackage[hyphens]{url}  
\usepackage{graphicx} 
\usepackage{natbib}  
\usepackage{caption} 
\usepackage{algorithm}
\usepackage{algorithmic}
\usepackage{amsmath,amssymb}
\usepackage{xspace}
\usepackage{wmx}
\usepackage{nameref}

\usepackage{newfloat}
\usepackage{listings}
\DeclareCaptionStyle{ruled}{labelfont=normalfont,labelsep=colon,strut=off} 
\floatstyle{ruled}
\newfloat{listing}{tb}{lst}{}
\floatname{listing}{Listing}
\title{\thistool{}: Branch-Aware Loop Invariant Inference with Large Language Models}
\author{
    Mingxiu Wang\textsuperscript{\rm 1},
    Jiawei Wang\textsuperscript{\rm 2},
    Xiao Cheng\textsuperscript{\rm 1}
}

\affiliations{
    \textsuperscript{\rm 1}Macquarie University\\
    \textsuperscript{\rm 2}University of New South Wales\\
    mingxiu.wang@students.mq.edu.au,
    jiawei.wang6@unsw.edu.au,
    xiao.cheng@mq.edu.au
}

\usepackage{bibentry}

\usepackage{xcolor}
\usepackage{tcolorbox} 

\definecolor{todocolor}{rgb}{0.9,0.1,0.1}

\begin{document}
\maketitle


\begin{abstract}
Loop invariants are fundamental for reasoning about the correctness of iterative algorithms. However, deriving suitable invariants remains a challenging and often manual task, particularly for complex programs. In this paper, we introduce \textsc{\thistool}, a branch-aware framework that integrates large language models (LLMs) to enhance the inference and verification of loop invariants. Our approach combines automated reasoning with branch-aware static program analysis to improve both precision and scalability. Specifically, unlike prior LLM-only guess-and-check methods, \textsc{\thistool} first verifies branch-sequence-level (path-level) clauses with SMT and then composes them into program-level invariants. We outline its key components, present preliminary results, and discuss future directions toward fully automated invariant discovery.

\end{abstract}

\section{Introduction and Motivation}
Loop structures are ubiquitous in programming and form the foundation of iterative reasoning. In traditional programs such as C, determining suitable invariants for \texttt{for} or \texttt{while} loops remains difficult, often relying on heuristics or manual insight. Similar iterative behaviors also appear in AI agents that repeatedly observe their surroundings, make decisions based on the current state, and iterate over actions or conditions until a specific goal is achieved~\cite{evertz2003loop, wu2022survey, mosqueira2023human}. Ensuring the correctness of these program-level loops is therefore crucial for the reliable functioning of such systems.  

A well-established approach to reasoning about loop correctness is through \emph{loop invariants}—properties that remain true before and after each loop iteration~\cite{furia2014loop, d2008survey}. Over the years, several methods have been proposed to automatically generate loop invariants.

Traditional approaches—including constraint solving~\cite{Coln2003LinearIG,8603013}, counterexample-guided refinement~\cite{clarke2000counterexample,bogomolov2017counterexample}, and Craig interpolation~\cite{jhala2006practical, mcmillan2010lazy} —encode invariants as logical constraints and iteratively refine them through symbolic reasoning. Although effective on small programs, these techniques require domain expertise to design templates or logical rules and often fail to scale to real-world software, where each new data structure demands specialized inference patterns.

Machine learning–based approaches leverage large code corpora to learn syntactic and semantic patterns and automatically generate invariants~\cite{garg2014ice,garg2016learning,nguyen2017counterexample,si2020code2inv}. These data-driven techniques generally offer better scalability than symbolic ones, as they reduce dependence on manually designed templates and rules by learning invariant patterns from data. 
However, they often suffer from issues such as overfitting, limited interpretability, and a strong dependence on the quality and diversity of the training data, leading to degraded performance on unseen programs or data structures.

More recently, large language model (LLM)–based approaches infer loop invariants directly from program code and natural-language prompts~\cite{wu2024llm,cao2025clause2inv,LIU2026103387}. These approaches typically follow a \emph{guess-and-check} paradigm: the LLM generates candidate invariants that are subsequently verified using SMT solvers or formal reasoning tools, with counterexamples fed back for refinement. 
Such methods are generally more flexible and adaptable, leveraging the extensive knowledge and contextual understanding embedded in LLMs.

However, despite their impressive capabilities, current LLMs still struggle to produce correct and precise invariants for programs with complex control flows, such as conditional branches. This limitation arises partly from the transformer architecture, which, while powerful in capturing global context, lacks explicit mechanisms for fine-grained structural reasoning and long-range code dependencies.

To address this challenge, we propose \thistool, a novel approach that combines the precision of program analysis that captures branch correlations with the generative and contextual strengths of LLMs. Our approach is designed to improve the precision and robustness of invariant generation for complex loops. This direction holds promise for advancing automated reasoning in program verification and synthesis tasks.

\section{Background and Related Work}

\subsection{Loop Invariants}

\noindent
A \textit{loop invariant} is a logical assertion $I$ that characterizes a property 
holding throughout the execution of a loop.  
In the Hoare triple
\[
\frac{
   P \Rightarrow I 
   \quad 
   \{\, I \wedge B \,\}\; S \;\{\, I \,\} 
   \quad 
   (I \wedge \lnot B) \Rightarrow Q
}{
   \{\, P \,\}\;\;\textbf{while } B \;\textbf{ do } S\;\;\{\, Q \,\}
}
\]

the components are defined as follows:
\begin{itemize}
    \item $P$: the \textbf{precondition}, describing the initial program state before the loop.
    \item $B$: the \textbf{loop guard}, a Boolean condition controlling loop execution.
    \item $S$: the \textbf{loop body}, the statement(s) executed when $B$ holds.
    \item $I$: the \textbf{loop invariant}, which must hold before and after each iteration.
    \item $Q$: the \textbf{postcondition}, describing the state after loop termination.
\end{itemize}

\noindent
The invariant $I$ must satisfy three fundamental conditions:
\begin{enumerate}
    \item \textbf{Initialization:} The precondition establishes the invariant ($P \Rightarrow I$).  
    \item \textbf{Preservation:} If $I$ holds and the guard $B$ is true, executing $S$ preserves $I$ ($\{ I \wedge B \} S \{ I \}$).  
    \item \textbf{Termination:} If $I$ holds and the guard $B$ is false, the postcondition follows ($(I \wedge \lnot B) \Rightarrow Q$).  
\end{enumerate}

\noindent
Thus, proving loop correctness reduces to finding an invariant $I$ that is 
\textit{established} by $P$, \textit{maintained} by $S$, and \textit{sufficient} to imply $Q$ when $B$ is false.

\subsection{State-of-the-Art Work}

Recent advances in invariant generation have explored leveraging large language models (LLMs) to automatically infer logical invariants. One of the most influential works in this direction is \textsc{Clause2Inv}~\cite{cao2025clause2inv}, which replaces the traditional \textit{guess–and–check} paradigm with a more systematic \textit{generate–combine–check} framework.

At a high level, \textsc{Clause2Inv} aims to generate an invariant $I$ that satisfies the three fundamental properties of loop correctness—\textit{Initialization}, \textit{Preservation}, and \textit{Termination}—by using an LLM to generate atomic clauses, the \texttt{Combinor} to combine them into candidate invariants, and an SMT solver to verify them. When any of these properties are violated, the system responds accordingly:

\begin{itemize}
    \item \textbf{If Initialization fails:} the generated invariant $I$ is not implied by the precondition $P$ ($P \nRightarrow I$).  
    In this case, the LLM is prompted to regenerate clauses that better align with the initial program state.

    \item \textbf{If Preservation fails:} executing the loop body $S$ under $I \wedge B$ does not maintain $I$ ($\{I \wedge B\} S \{\lnot I\}$).  
    Here, an SMT solver identifies counterexamples—specific program states where $I$ is broken.  
    These counterexamples are stored in a counterexample set (\texttt{ceSet}) and fed back to the LLM to refine its next generation.

    \item \textbf{If Termination fails:} the invariant $I$ is too weak to imply the postcondition $Q$ once the loop guard $B$ becomes false ($(I \wedge \lnot B) \nRightarrow Q$).  
    In this scenario, the system instructs the LLM to produce stronger invariants that can ensure the desired post-state.
\end{itemize}

To support this process, \textsc{Clause2Inv} first generates \textit{atomic clauses}—simple expressions such as $(a > 1)$ that exclude logical connectives like \texttt{\&\&} and \texttt{||}.  
These clauses are stored in an \texttt{exprStore}, and a dedicated \texttt{Combinor} component systematically combines them into more complex candidate invariants.  
Each candidate is formally verified using an SMT solver. Verified and sufficiently strong invariants are accepted; otherwise, the feedback loop described above is triggered.  
This iterative process continues until a valid invariant is found or the computation times out.

In this work, we adopt \textsc{Clause2Inv}~\cite{cao2025clause2inv} as our baseline and extend it by incorporating structural information from the program’s control-flow graph.  
By integrating control-flow relationships into the generation stage, our method guides the LLM toward producing invariants that are not only syntactically correct but also semantically aligned with the program’s execution structure.

\section{Method}
\label{sec:method}

In this section, we first provide an overview of the \thistool{} framework, and then describe in detail how the framework is implemented.
\begin{figure*}[t]
    \centering
    \includegraphics[width=1.00\textwidth]{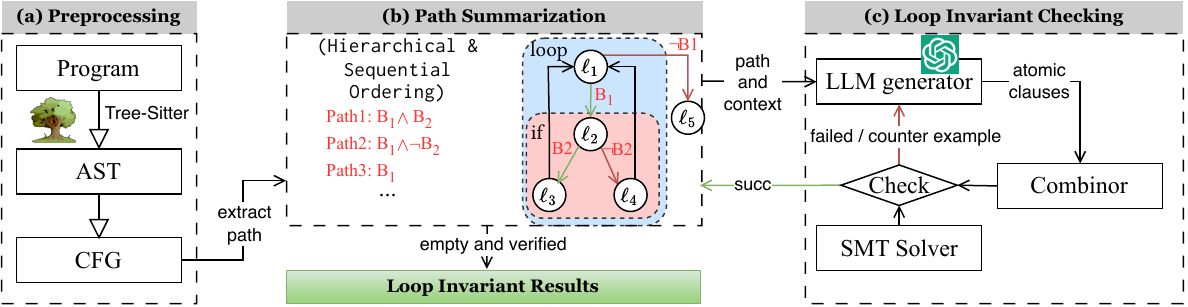}
    \caption{Overview of \thistool{}.}
    \label{fig:method_overview}
\end{figure*}
\subsection{Overview}

Figure~\ref{fig:method_overview} shows the framework overview of \thistool{} comprising the following three major phases:

\textbf{(a) Preprocessing.} Our approach begins by parsing the source code into an Abstract Syntax Tree (AST) using \textsc{Tree-sitter}~\cite{max_brunsfeld_2025_17180150}, which is then transformed into a Control Flow Graph (CFG) capturing both structural and execution relationships within the program. The CFG is traversed to extract all execution paths that encode branch sequences according to the program’s hierarchical and sequential structure (see Section~\nameref{sec:PathStore}).

\textbf{(b) Path Summarization.} Each path is independently analyzed by the LLM to generate clause-level invariants. Paths corresponding to inner or earlier control structures, such as the innermost loops or the first encountered branches, are assigned higher priority to ensure that their invariants are generated and verified first. 
Verified invariants are recorded as summaries for subsequent paths, allowing the LLM to leverage contextual knowledge during reasoning. Because each path corresponds to only a small program fragment, this path-based summarization improves reasoning accuracy and scalability.

\textbf{(c) Loop Invariant Checking.} After all paths are processed, the collected invariants and the complete source code are provided to the LLM to infer refined, program-level invariants. These candidates are combined and verified through an SMT solver. If verification fails, counterexamples are fed back to the LLM to guide correction—either by strengthening weak invariants or revising incorrect ones. This iterative \textit{generate–verify–refine} process continues until valid invariants are obtained or the computation times out.


\begin{algorithm}[t]
\caption{FindAllPaths in a Control Flow Graph (CFG)}
\label{alg:findallpaths}
\begin{algorithmic}[1]
\REQUIRE Control Flow Graph $CFG$ with entry node $CFG.entry$
\ENSURE List of all execution paths $Paths$
\STATE $Paths \gets \varnothing$
\STATE $node \gets CFG.entry$
\WHILE{$node \neq$ null}
    \IF{$node.out\_degree > 1$}
        \IF{$node$ is a loop header}
            \STATE $loopCFG \gets$ \textsc{ExtractLoopCFG}($node$)
            \STATE $loopPaths \gets$ \textsc{FindAllPaths}($loopCFG$)
            \STATE $Paths \gets Paths \cup loopPaths$
        \ELSE 
            \STATE $trueCFG \gets$ \textsc{GetTrueCFG}($node$)
            \STATE $falseCFG \gets$ \textsc{GetFalseCFG}($node$)
            \STATE $truePaths \gets$ \textsc{FindAllPaths}($trueCFG$)
            \STATE $falsePaths \gets$ \textsc{FindAllPaths}($falseCFG$)
            \STATE $branchPaths \gets truePaths \cup falsePaths$
            \STATE $Paths \gets Paths \cup branchPaths$
        \ENDIF
    \ELSIF{$node.out\_degree = 1$}
        \STATE \textsc{AddToContext}($node$)
    \ENDIF
    \STATE $node \gets node.successor$
\ENDWHILE
\RETURN $Paths$
\end{algorithmic}
\end{algorithm}


\subsection{Path Extraction Algorithm}
\label{sec:PathStore}
To systematically extract all execution paths from a program, we propose Algorithm~\ref{alg:findallpaths}, which operates on the program's Control Flow Graph (CFG).  
The algorithm traverses the CFG starting from the entry node and explores each branch and loop to enumerate possible execution paths. For nodes with multiple outgoing edges, it distinguishes between loop headers and general branching points.  

For loop headers, the algorithm identifies the sub-CFG corresponding to the loop body (which has already been generated during CFG construction from the AST) and recursively computes all possible execution paths within the loop, appending them to the overall path list. For general branching nodes, it recursively computes and combines paths along both the true and false successors. Nodes with a single successor are added to the context, which is subsequently used as input to the LLM. The algorithm terminates when all reachable paths from the entry node have been explored.


\subsection{Hierarchical Path Summarization with Verified Invariants}

Algorithm~\ref{alg:hierarchical_abstraction} refines the set of execution paths obtained from Algorithm~\ref{alg:findallpaths} by summarizing compound paths—such as loop bodies and conditional branches—into verified invariants. Its goal is to replace complex control-flow segments with their logical summaries, enabling reasoning about correctness at a higher level of abstraction.

Given an input list of execution paths, the algorithm processes each path iteratively. For loops, it identifies the sub-paths within the loop body, recursively computes their invariants, and replaces the loop segment with a loop invariant verified using an SMT solver. Similarly, for conditional branches, invariants are computed for both the true and false branches, and the corresponding paths are replaced with their logical disjunction.

A global precondition is maintained to accumulate constraints across paths, while separate stacks record active loops and branch structures. Together, these elements form the context provided to the LLM, enabling it to generate stronger and more accurate invariants. In Algorithm~\ref{alg:hierarchical_abstraction}, this context includes the global precondition, the loop stack, the branch stack, and other relevant program information.


\begin{algorithm}[t]
\caption{Hierarchical Path Summarization with Verified Invariants}
\label{alg:hierarchical_abstraction}
\begin{algorithmic}[1]
\REQUIRE List of execution paths $Paths$
\ENSURE Verified invariant summarizing all paths
\STATE $pre\_cond \gets \varnothing$ \COMMENT{Global precondition}
\STATE $loop\_stack \gets$ empty stack
\STATE $branch\_stack \gets$ empty stack
\FOR{each $path \in Paths$}
    \IF{$path$ contains a loop}
        \STATE Push loop context to $loop\_stack$
        \STATE $loop\_body \gets$ extract subpaths within the loop
        \STATE $loop\_inv\gets$HierarchSummarize($loop\_body$)
        \STATE $pre\_cond \gets pre\_cond \wedge loop\_inv$
        \STATE Pop loop context from $loop\_stack$
    \ELSIF{$path$ contains a conditional branch}
        \STATE Push branch context to $branch\_stack$
        \STATE $true\_branch, false\_branch \gets$ extract branches
        \STATE $true\_inv \gets$ HierarchSummarize($true\_branch$)
        \STATE $false\_inv \gets$ HierarchSummarize($false\_branch$)
        \STATE $pre\_cond \gets pre\_cond \wedge (true\_inv \vee false\_inv)$
        \STATE Pop branch context from $branch\_stack$
    \ELSE
        \STATE $path\_inv \gets$ ComputeInvariant($path$)
        \STATE $pre\_cond \gets pre\_cond \wedge path\_inv$
    \ENDIF
\ENDFOR
\STATE $context \gets (loop\_stack, branch\_stack, pre\_cond)$
\STATE $invariants \gets$ getInvariantsFromLLM($context$)
\RETURN $invariants$
\end{algorithmic}
\end{algorithm}

\section{Early Results and Future Plan}

We conducted preliminary experiments on a subset of benchmark programs to evaluate the effectiveness of our branch-aware loop invariant generation approach in addressing the following research question (RQ): ``Can \thistool{} perform better than existing LLM-based loop invariant generation methods in terms of accuracy?'' We also summarize our future plan to further enhance and validate our approach.

\subsection{Preliminary Results}

We selected benchmark programs from the widely used SV-COMP suite, which includes a diverse set of programs with varying complexity and control-flow structures, such as parallel loops and nested loops. First, we implemented state-of-the-art LLM-based loop invariant generation tools, such as Clause2Inv and Lam4Inv, to evaluate whether they could generate correct invariants for these programs. The results show that these tools struggle with programs that have complex control flow, often failing to generate correct invariants within a reasonable time frame.

Next, we extracted the different execution paths from these programs and fed them into the LLMs to generate path-specific invariants. The correctness of each generated invariant was then verified using an SMT solver. We replaced the corresponding paths with these verified invariants and repeated the process until all paths were processed. Finally, we combined all the generated invariants along with the source code as input to the LLM to produce the final invariants for the entire program, which were also checked using the SMT solver.

The preliminary results are promising. Our structure-aware approach successfully generated correct invariants for programs that were previously challenging for existing methods. In particular, we observed a significant improvement in accuracy, with our method achieving a higher success rate in generating correct invariants. This improvement is attributed to the focused reasoning provided by path-specific analysis, which allows the LLM to better understand the context and constraints of each execution path, concentrating on a small portion of the program at a time and thereby reducing the complexity of the reasoning task.

\subsection{Future Directions}

In summary, our future work focuses on expanding the dataset, automating the end-to-end workflow, and improving the effectiveness and scalability of our branch-aware loop invariant generation approach.

First, we plan to extract a more diverse set of benchmark programs from the SV-COMP suite, focusing on programs with complex control structures, intricate arithmetic operations, and array manipulations. This will allow us to better evaluate the generalization capabilities of our approach across different program types and identify areas for improvement.

Second, we aim to implement the algorithms described in Section~\nameref{sec:method} to achieve a fully automated, end-to-end workflow. By integrating path extraction, hierarchical path summarization, LLM-based loop invariant generation, and SMT-based verification, we can minimize manual intervention and streamline the entire loop invariant generation process.

Finally, we plan to conduct a comprehensive evaluation of our approach against state-of-the-art LLM-based loop invariant generation tools on large-scale and industrial benchmarks. This evaluation will consider accuracy, efficiency, and scalability. Additionally, we will explore enhancements such as:
\begin{itemize}
    \item Context-aware LLM prompting strategies to improve invariant precision.
    \item Hybrid techniques combining symbolic reasoning with LLM predictions.
    \item Extending the approach to handle concurrency, recursion, and other advanced constructs.
\end{itemize}

Through these efforts, we aim to develop a robust, scalable, and practical branch-aware loop invariant generation framework that outperforms existing methods.

\section{Conclusion}

This paper examined the limitations of current LLM-based invariant generation tools in handling programs with complex control-flow, such as loops and nested branches. To address these challenges, we proposed \thistool{}, a path-based framework that generates path-specific invariants verified with SMT solvers. By decomposing programs into individual execution paths, \thistool{} simplifies the reasoning task for the LLM and improves the correctness of generated invariants.

Preliminary results on benchmark programs show that \thistool{} successfully generates correct invariants for programs that are challenging for existing methods, demonstrating both higher accuracy and promising efficiency. Future work includes expanding the benchmark set, automating the end-to-end workflow, and exploring context-aware and hybrid reasoning strategies to further enhance performance and scalability.

\bibliography{aaai2026}
\end{document}